\documentclass[aip,rsi,reprint,graphicx,twocolumn]{revtex4}

\usepackage{epsfig}
\usepackage{graphicx}
\usepackage{dcolumn}
\usepackage{bm}

\usepackage[paperwidth=225mm,paperheight=289mm,centering,hmargin=1.5cm,vmargin=1.5cm]{geometry}

\begin{document}


\title{Probabilistic Optically-Selective Single-molecule Imaging Based Localization Encoded (POSSIBLE)  Microscopy for Ultra-superresolution Imaging }

\author{ Partha Pratim Mondal }
\email[Corresponding author: partha@iap.iisc.ernet.in]{}
\affiliation{%
Nanobioimaging Lab, Department of Instrumentation $\&$ Applied Physics, Indian Institute of Science, Bangalore 560012, INDIA \\
}%
\date{\today}
        
\begin{abstract}
	
To be able to resolve molecular-clusters it is crucial to access vital informations (such as, molecule density and cluster-size) that are key to understand disease progression and the underlying mechanism. Traditional single-molecule localization microscopy (SMLM) techniques use molecules of variable sizes (as determined by its localization precisions (LPs)) to reconstruct super-resolution map. This results in an image with overlapping and superimposing PSFs (due to a wide size-spectrum of single molecules) that degrade image resolution. Ideally it should be possible to identify the brightest molecules (also termed as, the {\bf{fortunate molecules}}) to reconstruct ultra-superresolution map, provided sufficient statistics is available from the recorded data. POSSIBLE microscopy explores this possibility by introducing narrow probability size-distribution of single molecules (narrow size-spectrum about a predefined mean-size). The reconstruction begins by presetting the mean and variance of the narrow distribution function (Gaussian function). Subsequently, the dataset is processed and single molecule filtering is carried out by the Gaussian distribution function to filter out unfortunate molecules. The fortunate molecules thus retained are then mapped to reconstruct ultra-superresolution map. In-principle, the POSSIBLE microscopy technique is capable of infinite resolution (resolution of the order of actual single molecule size) provided enough fortunate molecules are experimentally detected. In short, bright molecules (with large emissivity) holds the key. Here, we demonstrate the POSSIBLE microscopy technique and reconstruct single molecule images with an average PSF sizes of $\sigma \pm \Delta\sigma = 15\pm 10 ~nm, ~30\pm 2 ~nm ~\& ~50\pm 2~nm$. Results show better-resolved Dendra2-HA clusters with large cluster-density in transfected NIH3T3 fibroblast cells as compared to the traditional SMLM techniques. \\        

\end{abstract}

\maketitle

Imaging molecular processes with true molecular-resolution is the ultimate goal of light microscopy. This is largely eluded due to the diffraction of light which sets a lower bound on the resolution-limit \cite{abbe}\cite{rayleigh}. However, recent microscopy techniques such as, STED \cite{hell1994}, fPALM \cite{hess2006}, STORM \cite{rust2006}, PALM \cite{betzig2006}, SIM \cite{sim2005} GSDIM \cite{fol2008} SOFI \cite{sofi2009}, PAINT \cite{paint1} \cite{paint2}, SMILE \cite{smileMRT} \cite{smileAPL}, MINFLUX \cite{minflux} and others techniques \cite{joerg} \cite{abu} \cite{mike} \cite{vermal} \cite{matt} \cite{ma} \cite{turcotte} \cite{valles} \cite{wang} have surpassed this limit \cite{abbe}\cite{rayleigh}. Still the above techniques are far-from truly molecular-scale resolution and thus are incapable of functional imaging with ultra-highresolution (preferably in the range single molecule lengthscales). Hence it becomes  imperative to develop technique that can explore sub-10 nm domain.  \\

Single molecule localization microscopy rely on the fact that each detected molecules can be localized within a certain distance about its centroid. This determines its size and PSF.  The localization precision (LP) is heavily dependent on the number of photons emitted and detected by the Camera. So, a low photon number count for a recorded molecule is most likely to produce large PSF leading to a poorly-resolved image and vice-versa. A better way to resolve single molecules would be to selectively choose the fortunate molecules that emits large number of photons, thereby reducing localization uncertainty and its size. Existing SMLM techniques (fPALM/PALM) have large variability in their sizes (or LPs) since the PSFs have large size-spectrum ranging from few tens to few hundred nanometers. So, these super-resolution techniques are most likely to produce images with large mean resolution. On the other hand, a image reconstructed using fortunate molecules (having small PSFs and narrow size distribution) would lead to better resolution. Such a map is useful since existing localization microscopy confront situations where the molecules exibit mixed emission properties (accounting for both weak and strong emissions) and incorporation of all these molecules to build-up a map result in compromised resolution. Hence, a better strategy would be to look for fortunate molecules during data acquisition and record enough molecules to reconstruct the image. In a multi-resolution setup, one can define multiple Gaussians to classify molecules into strong, moderate and weak emitters based on their photon count. This helps to built a multi-resolution map of jumbled clusters which otherwise were poorely resolved by the existing SMLM techniques. \\

Here, we propose a new localization technique (termed as, POSSIBLE microscopy) which is capable of producing multiresolution single molecules maps. In principle, the technique allow actual-molecular resolution (in the range of few-nanometers) provided the detection of bright photoactivable probes is achieved. To investigate viral infection due to influenza, we studied Hemagglutinin (HA) expressed in NIH3T3 fibroblast cells \cite{white}, its distribution and number density. These parameters are critical for viral replication and subsequent release of bud particles from infected cells \cite{chen2007}. Although existing super-resolution microscopy techniques show the presence of clusters during infection \cite{sam2007} \cite{sam2013}, it has never allowed in-depth study of molecular clusters due to limited resolution. This requires immediate attention since intra-cluster resolution and packing-fraction are essential to determine the rate of infection. Our results reveal the existence of dense molecular sub-clusters (within large clusters) post 24 Hrs of infection in HA transfected NIH3T3 cells. These informations are vital for developing antiviral drugs for disrupting HA assembly in order to subside the infection. \\       

In this communication, we propose a technique that can extend the resolution obtained by traditional localization microscopy. The proposed method explores for fortunate molecules that can be better localized (small average size and narrow variance) using a probability distribution for molecular size (or equivalently LP). This technique can be used to explore resolution of the order of actual molecule dimension (sub-$10~nm$ resolution) for live biological specimens. \\

\begin{figure*}
	\begin{center}
		\includegraphics[width=19.5cm, angle=0]{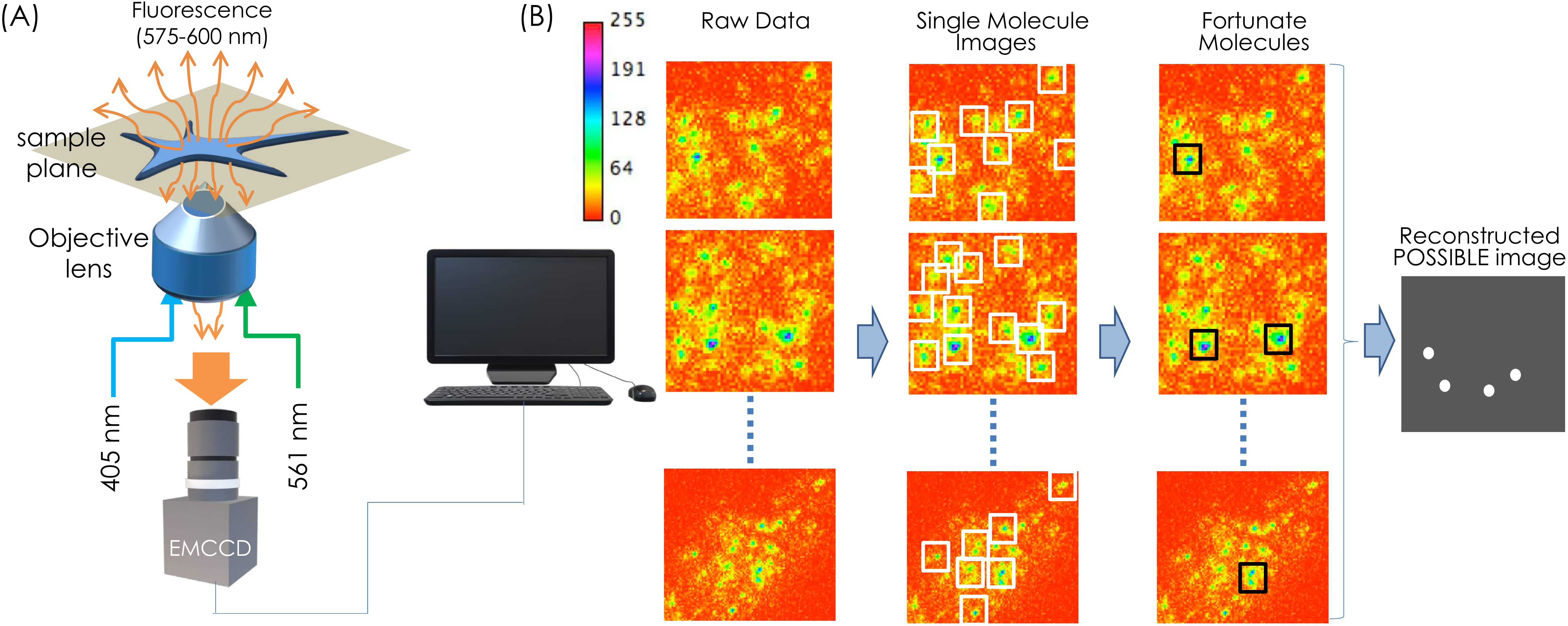}
		\caption{POSSIBLE microscopy, (A) The schematic diagram of optical setup, (B) Computational processes for determining the single molecules from the raw data after background substraction and thresholding. This is followed by Gaussian filtering to retain the fortunate molecules for reconstructing POSSIBLE image. }
	\end{center}
\end{figure*}

\section{Results}

POSSIBLE microscopy has the ability to discern details within molecular systems such as molecular clusters. In this study, we employ POSSIBLE microscopy  to analyze HA clusters and compare it with the existing SMLM techniques.

\subsection{POSSIBLE Multiresolution Microscopy }

POSSIBLE microscopy looks for fortunate molecules that falls in the Gaussian distribution with predefined LP mean and variance. In general, localization microscopy techniques utilizes all the recorded single molecule PSFs with a well-defined localization precision that determines their size. So, the reconstructed map is a collection of single molecules / PSFs with localization precision varing from few tens to hundred nanometers. This results in blurred clusters due to overlapping PSFs (corresponding to single molecules) of varying sizes. To overcome the resolution obstructed by large PSFs, it may be possible to construct maps with fortunate molecules with relatively small PSFs. \\   

In this article, we report a new multi-resolution optical microscopy technique to resolve molecular clusters (see, Fig. 1). A possible yet effective way to address this would be to collect fortunate molecules from a large dataset. All the molecules have to pass through the probabilistic Gaussian filter test and depending upon their respective probability, they are selected to the set that represent the reconstructed image with desired resolution (calculated based on PSF size and size-variance). \\

\begin{figure*}
	\begin{center}
		\includegraphics[width=19cm, angle=0]{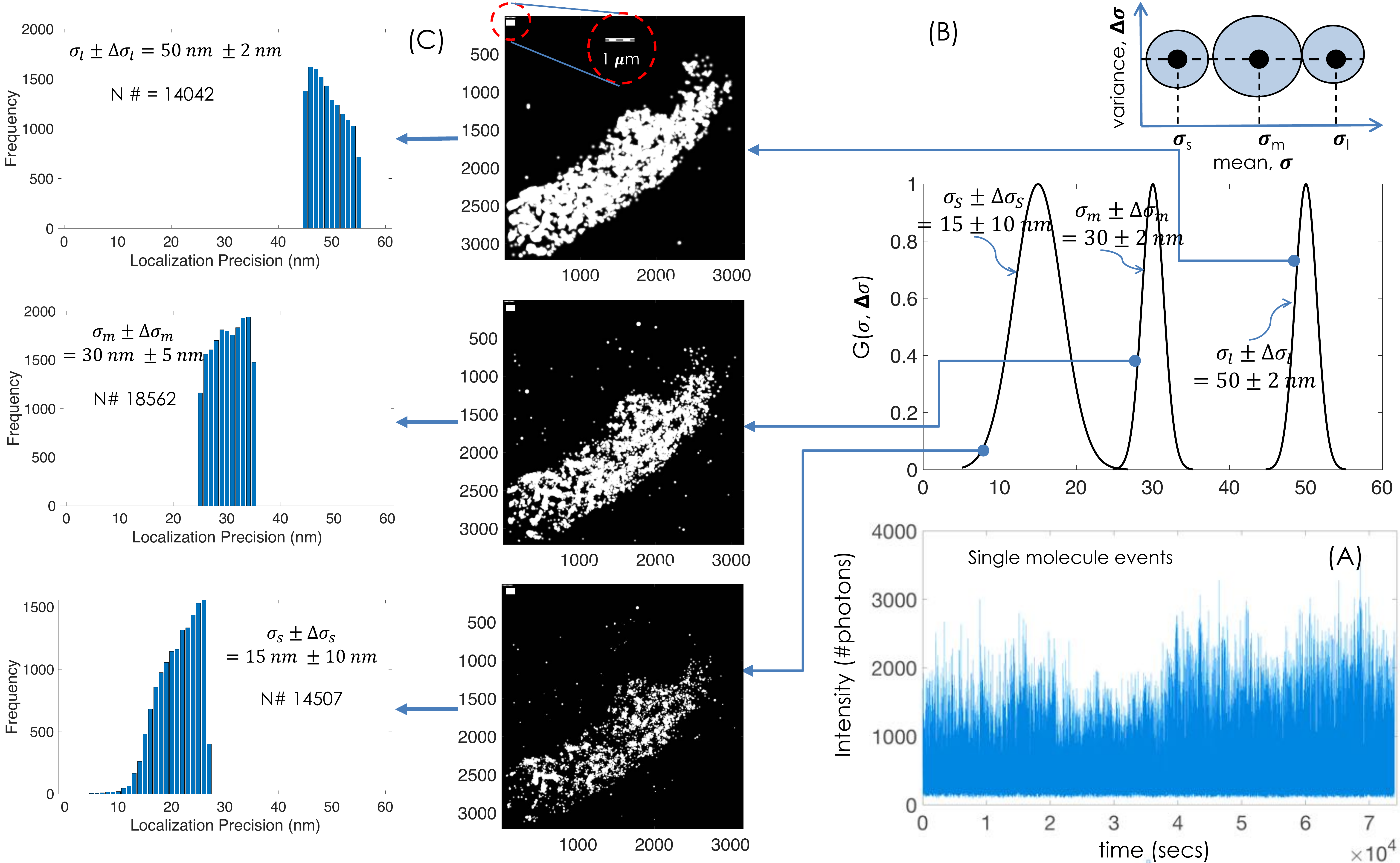}
		\caption{ Multiresolution ultra-superresolution imaging using POSSIBLE microscopy. (A) The histogram of single molecule images as recorded by the EMCCD detector. (B) Three different Gaussian filters ($G(\sigma_s , \Delta\sigma_s)$, $G(\sigma_m , \Delta\sigma_m)$ and $G(\sigma_l , \Delta\sigma_l)$) for reconstructing 3 distinct POSSIBLE images. Schematic of $(\sigma, \Delta\sigma)$ space is also shown, (C) Reconstructed images with the corresponding localization precision histogram for highly-resolved molecules (corresponding to $G(\sigma_s , \Delta\sigma_s)$), moderately-resolved molecules ($G(\sigma_m , \Delta\sigma_m)$) and poorly-resolved molecules ($G(\sigma_s , \Delta\sigma_s)$) respectively. }
	\end{center}
\end{figure*} 

\begin{figure*}
	\begin{center}
		\includegraphics[width=18cm, angle=0]{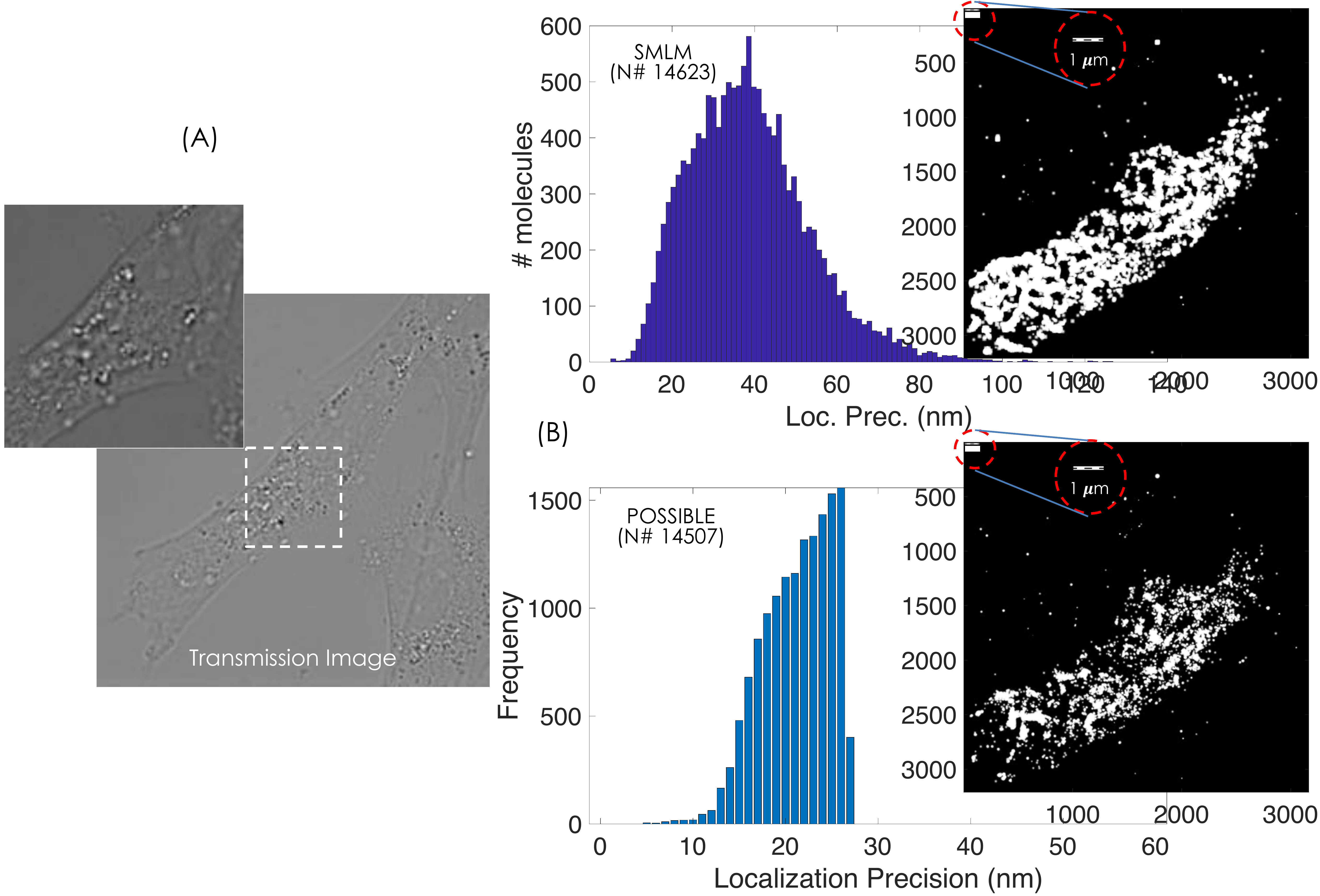}
		\caption{Comparison of POSSIBLE microscopy with SMLM (PALM/fPALM). (A) The transmission image of the cell. (B) Both SMLM and POSSIBLE reconstructed images are shown along with the histogram for number of molecules / frequency vs. LP. The comparison is carried out with almost equal number of molecules ($N\#$) to represent molecular clusters in the cell. }
	\end{center}
\end{figure*}

Fig. 2 show images obtained using multiple Gaussian filters. The single molecule events as recorded by the detector is shown in Fig. 2A. Three different Gaussian distribution were considered based on their size (or equivalently their LP): poorly-resolved ($G(\sigma_{lp} ~,~ \Delta\sigma_{lp})=G(50nm~,~2 nm)$), moderately-resolved ($G(30nm~,~2 nm)$) and highly-resolved ($G(15nm~,~10 nm)$) as shown in Fig.2B. A cartoon representing $\sigma$ vs $\Delta\sigma$ along with their LP means is also shown. The reconstructed images along with the LP bar-plot are shown in Fig. 2C. The $(\sigma ~, ~\Delta\sigma)$ were chosen such that we get nearly equal number of molecules in each distribution to enable comparison. The corresponding histograms for localization precision are also shown. The histogram represent frequency of occurrance of single molecule with specific $(\sigma, \Delta\sigma)$. The number of molecules ($N\#$) for $G(\sigma_{l} ~,~ \Delta\sigma_{l})$, $G(\sigma_{m} ~,~ \Delta\sigma_{m})$ and $G(\sigma_{s} ~,~ \Delta\sigma_{s})$ are 14042, 18562 and 14507, respectively. Obviously, this shows that the recorded raw data has relatively large number of moderately-resolved molecules. In fact, several experiments show low statistics for highly-resolved molecules indicating significantly lower occurrance of molecules that emit large number of photons. Visually, the reconstructed super-resolution map reveals that the clusters are better-resolved for $G(\sigma_{s} ~,~ \Delta\sigma_{s})=G(15nm~,~10 nm)$ as compared to $G(50nm~,~2 nm)$ and $G(30nm~,~2 nm)$. This is due to the fact that map generated by $G(\sigma_{s} ~,~ \Delta\sigma_{s})$ has smaller PSFs compared to those for $G(\sigma_{m} ~,~ \Delta\sigma_{m})$ and $G(\sigma_{l} ~,~ \Delta\sigma_{l})$. So reconstruction of super-resolution image with fortunate molecules (falling in the distribution, $G(15nm~,~10 nm)$) result in ultra-high resolution (better than traditional SMLM image).  \\

\subsection{Comparison with SMLM Microscopy}

To demonstrate the advantages of POSSIBLE microscopy, a comparison with the state-of-the-art SMLM (fPALM/PALM) was carried out. To ascertain fare comparison, we have ensured that same number of molecules were used for reconstructing image. Fig. 3A shows the transmission image of the HA transfected NIH3T3 fibroblast cell and the region being imaged (white dotted square). Fig. 3B shows POSSIBLE resonstructed image built from highly-resolved single molecules ( i.e., $G(15~nm, ~10~nm)$). A total of 14507 molecules out of 173448 recorded molecules (from 10,000 frames) have passed the Gaussian filtering process and found qualified to represent the reconstructed image. The second part of Fig.3B shows super-resolution image obtained using traditional SMLM microscopy (with actual molecule size or LP). The image is reconstructed using 14623 single molecules. Visual inspection show better-resolved structures for POSSIBLE microscopy. This is due to the fact that single molecules used to reconstruct POSSIBLE microscopy has a mean size of $\approx 15~nm$ with a narrow size-spectrum ($\approx ~20 ~nm$) whereas, the SMLM reconstructed image has a relatively large mean size of $\approx 40 ~nm$ with a wide size-spectrum ($>100 ~nm$). Thus restricting molecules with large sizes (corresponding to large LPs) to participate in the image reconstruction process has resulted in better resolved molecular structures. It may be noted that the number of molecules with size $< 10~nm$ are significantly few indicating the inability of POSSIBLE microscopy to reconstruct images with mean size $<10~nm$ pertaining to the present dataset. However, a better dataset with ultra-bright molecules is capable of resolving single-molecule lengthscales. So, POSSIBLE microscopy can in-principle provide unlimited resolution (spatial resolution of the actual size of single molecule), provided experimental detection of enough fortunate molecules is achieved.  \\

To visually illustrate the resolving capabilities of the POSSIBLE and SMLM microscopy, we have chosen two different HA clusters (marked as R1 and R2 by the blue and red squares respectively) in the reconstructed map, and carried out 3D surface-plots (Fig. 4). The zoomed version of the clusters along with 3D surface plots are also shown in Fig. 4. SMLM reconstructed cluster show overlapping PSFs, whereas clusters reconstructed by POSSIBLE microscopy reveals relatively sparse PSFs. This is due to the involvement of small-sized PSFs (range, $5 - 25~nm$ corresponding to $G(15~nm~,~ 10~nm)$) of fortunate molecules in the cluster formation that gives resolution boost to POSSIBLE microscopy. On the other hand, SMLM builts the reconstruction map with PSFs having large size-spectrum ($10~nm - 100~nm$). The dark circles show appearance of sub-clusters within large clusters by POSSIBLE microscopy which are otherwise not resolved by SMLM. This suggests that POSSIBLE microscopy is better suited for resolving molecular clusters. \\

\subsection{HA Cluster Analysis}

Clustering during influenza infection is critical to access its current state and progression rate. The reconstructed data (see, Fig.2) shows clustering of single molecules 24 Hrs post-transfection. Many factors contribute to the clustering of HA molecules and the consequent increase in cluster-size. In addition, the determination of number of HA molecules in a cluster is critical and has a strong influence on the rate of infection. These studies require optical microscopes that can resolve molecular clusters. \\

Our study with NIH3T3 fibroblast cell lines using photoactivable HA molecules show high-resolution image of HA distribution. In order to elucidate the potential of POSSIBLE microscopy, we study cluster size, the number of HA molecules in the clusters and its distribution. Raw data of single molecules is analyzed  post-reconstruction followed by Matlab based rendering algorithms \cite{cleanAIP,smileMRT, smileAPL}. The clustering process begins by obtaining the location information of rendered single molecules (x and y positions) from the master data-matrix that contains other information such as, localization precision, number of detected photons per molecule and others. The location information is fed to the developed clustering algorithm primarily based on k-means clustering (see, Matlab scripts). Fig. 5 shows multiple clusters along with their centroids for POSSIBLE  microscopy ($G(15~nm~, ~10~nm)$). The clusters are color-coded with their centroids marked by $'*'$. Since the clusters occur in all shapes, it is difficult to uniformly determine its size assuming isotropic spread of the cluster. So, we preferred to study cluster area rather than cluster size in the present work. \\

Figure 5 shows K-means clustered data along with cluster-density analysis (the number of molecules ($\#N$) versus cluster number (Cluster $\#$) ) (see inset in Fig. 5). One can readily see that some clusters in SMLM reconstructed map looks elongated, however POSSIBLE reconstructed image show sub-clusters that are more compact and alike. Moreover, we observed a sharp decrease in large-sized clusters (of area $\approx 6 ~\mu m^2$) in POSSIBLE reconstructed map ($\sigma_{lp} \pm \Delta\sigma_{lp} = 15\pm 10 ~nm $) as compared to SMLM reconstructed map. Furthermore, the dominance of mid-size clusters (of area $1-4 ~\mu m^2$) with the appearance of sizeable large clusters (of area $\approx 6 ~\mu m^2$) is quite evident (data not shown). This represents critical stage of clustering process post 24 Hrs since, the transfected cells were processed and fixed after 24 Hrs. This observation is significant because large assembly of HA are know to occur much beyond 24 Hrs of infection. \\


K-means clustering suggests a total of 26 and 23 clusters, (including sparse ones) respectively for POSSIBLE and SMLM. The size of the clusters have ranged from $0.5 ~ - ~ 6 ~\mu m^2$ and the number of HA molecules per cluster range from few hundred to few thousand (data not shown). Cluster-density analysis indicates the presence of few highly-dense clusters (about $10\%$) and the rest $90\%$ clusters are of low density ($<600$ molecules). Other analysis corroborating the above observation is the cluster spread-density (total HA molecules per cluster area-spread ) that indicates a similar trend. Overall, the studies suggests strong clustering, whereas dense clusters are better enunciated by POSSIBLE microscopy. The study also indicates the presence of medium-sized clusters. This indicates the presence of clusters with densely-packed HA molecules. We attribute this to the ability of POSSIBLE microscopy that employs relatively small-sized PSFs for the reconstruction of super-resolution map. The study shows the appearance of dense HA clusters post 24 Hrs of transfection. This is in consistence with the observation that dense packing / comncentrations of HA (influenza viral membrane protein hemagglutinin (HA)) is essential for infectivity \cite{rein, gott, bry, sam2013}. \\

\begin{figure*}
	\begin{center}
			\includegraphics[width=19cm, angle=0]{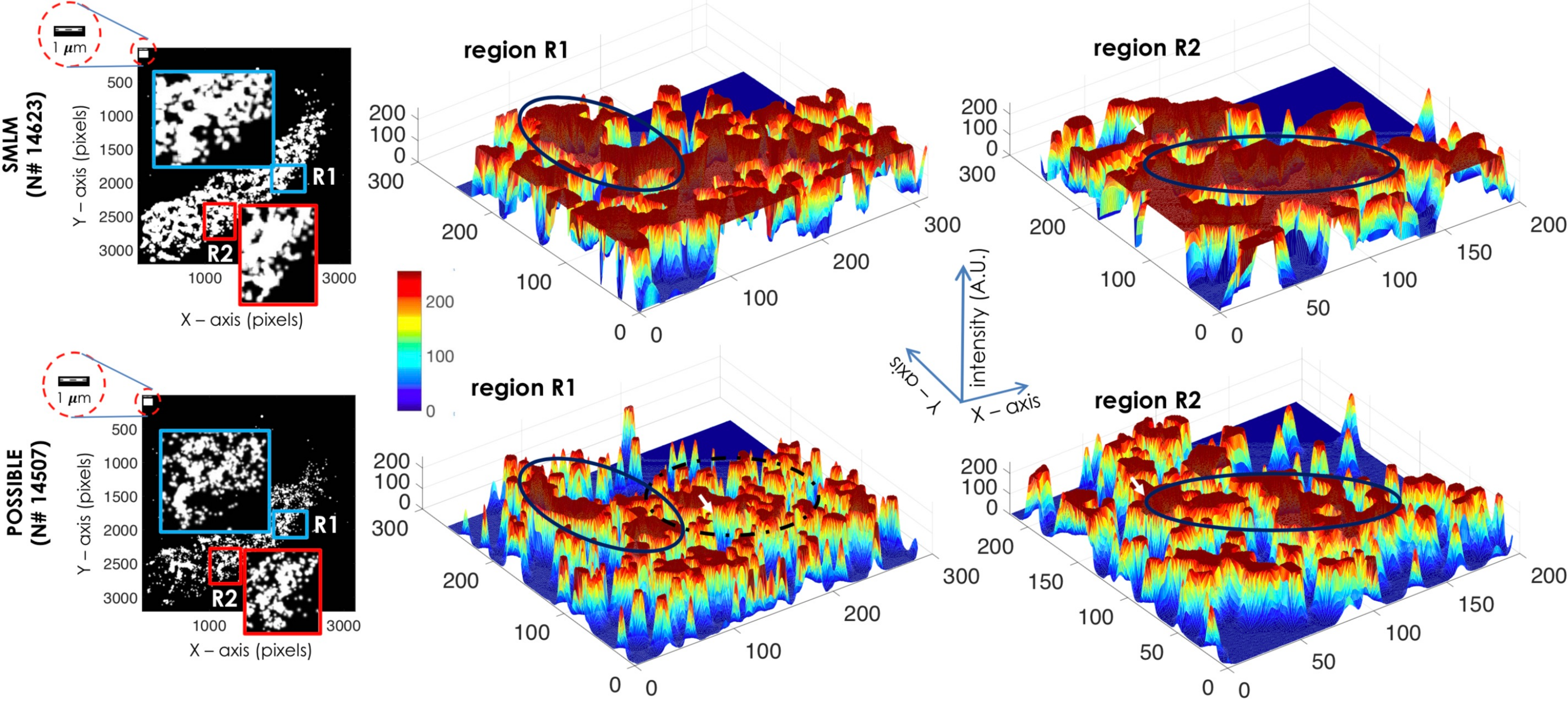}
		\caption{Cluster Visualization. (A) SMLM and POSSIBLE reconstructed images along with the zoomed version of two different clusters (marked by blue and red rectangles). (B) 3D surface plot of two chosen clusters (marked by R1 and R2) where the marked black ellipses show sub-clusters within large clusters. Visually, the resolving ability of POSSIBLE microscopy is quite apparent. }
	\end{center}
\end{figure*} 

\begin{figure*}
	\begin{center}
		\includegraphics[width=19.5cm, angle=0]{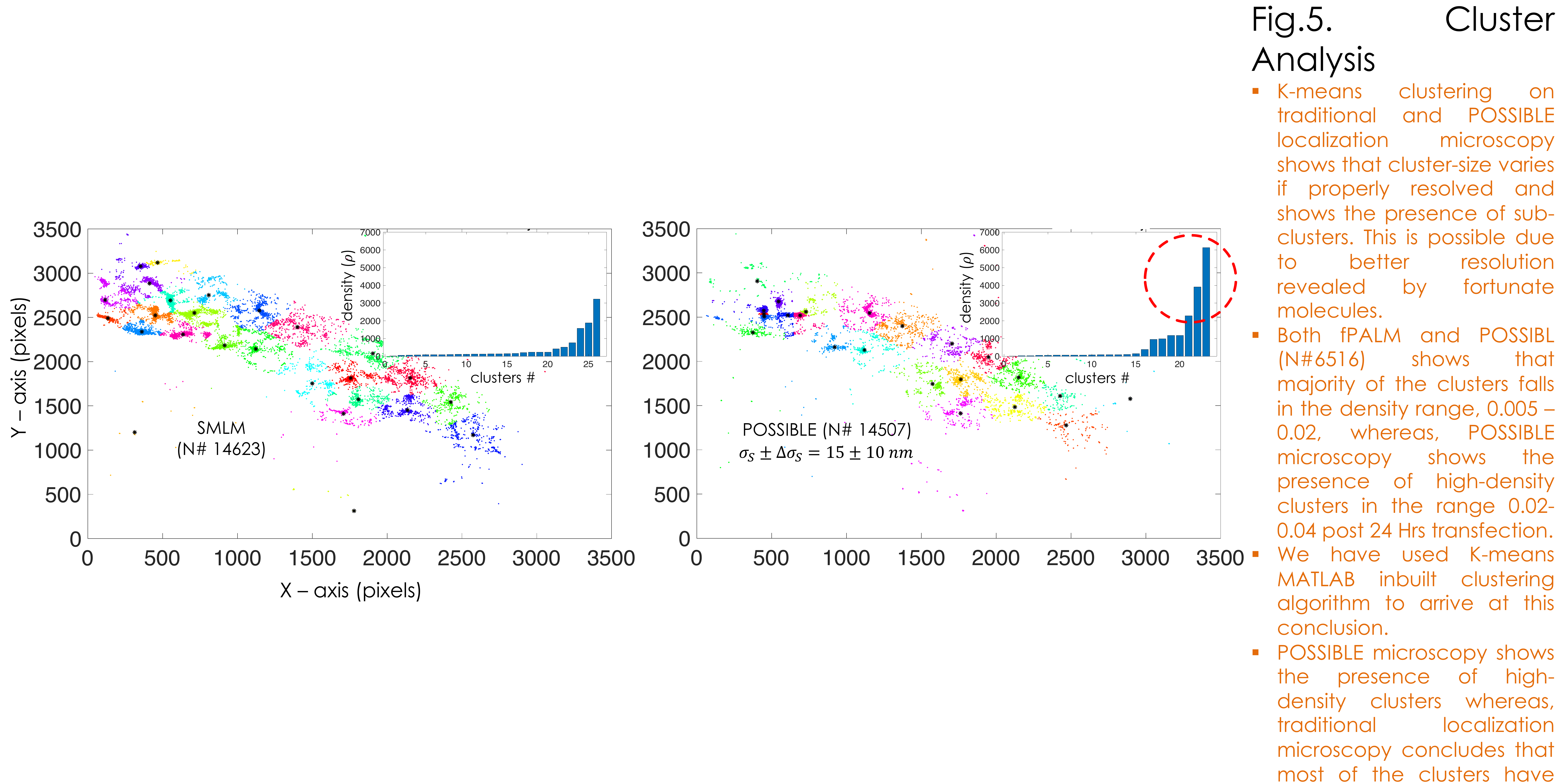}
		\caption{Cluster analysis using K-means clustering algorithm for SMLM (PALM/fPALM) and POSSIBLE microscopy (with filter, $G(\sigma_s , \Delta\sigma_s)$). Cluster density analysis are also shown in the inset. The red circle indicates high-density  clusters for POSSIBLE reconstructed images ($G(\sigma_s , \Delta\sigma_s)$) which is missing for SMLM.  }
	\end{center}
\end{figure*}

\section{Methods and Protocols}

\subsection{POSSIBLE Microscopy: Optical Setup}

POSSIBLE microscopy is built upon the standard localization microscopy (SMILE / fPALM)  \cite{smileAPL} \cite{hess2006}. The schematic diagram depicting the key components of the microscope and associated computational processes are shown in Fig.1. A laser light of wavelength, $\lambda_{acv} =405~nm$ (Oxxius, France) was used to randomly activate the molecules and a second laser of wavelength, $\lambda_{exc} = 561~nm$ (Oxxius, France) was used for exciting the molecules. We used a high numerical aperture (NA) objective (Olympus 1.3NA, 100X) for both illuminating the sample and collecting the fluorescence photons from single molecules. The molecules (Dendra2HA) have an emission peak at $573~nm$. A filter-cube containing dichroic filter and bandpass filter (from Shemrock, USA) was used to direct the light to specimen and filter-out fluorescence. An additional set of filters were employed in the detection path (Notch filters, 405 nm and 561 nm from Shemrock, USA) to block the back-reflected laser lights. Other optical components such as, mirrors, beam-splitters and opto-mechanical components were purchased from Thorlabs, Newton, NJ. The images were recorded using an Andor 897 iXon Ultra camera (Andor Technology, UK). \\

During experimentation, a large number of images (about 10000) were recorded and was followed by computational processing. The raw data was recorded by the EMCCD detector and subjected to analysis for identifying single molecules (see, white boxes in Fig. 1B). Subsequently, important parameters (such as, detected photons per molecule, centroid, molecule position, variance and others) were extracted from the analysis. A range for the size of molecules was preset to remove random noise (from the acquited images) and pixel-to-photon conversion was carried-out. The extracted data was then subjected to Gaussian filter (using Gaussian probability distribution function) to filter-out unfortunate molecules and retain the fortunate ones. The fortunate molecules are represented by black square boxes as shown in Fig.1B. The final step involves the reconstruction of single molecule image (termed as, POSSIBLE ultra-superresolution image) from the detected fortunate molecules. \\

\subsection{Gaussian Filter and Fortunate Molecules}

The key to multi-resolution POSSIBLE microscopy is the identification of fortunate molecules. A Gaussian probabilistic function is utilized to determine the size-spectrum of molecules needed to reconstruct the map. In general, molecules are defined by their centroid position and localization precision (say, $\sigma$) which determines their size \cite{thompson}. Localization precision is used to determine the position uncertainty around the centroid of molecule that fixes its size. For convenience, we define three sets of molecules with mean LP ($\sigma$) and its variance $\Delta\sigma$. Accordingly, we choose to work with Gaussian distribution. Three Gaussian distributions in the $(\sigma, ~\Delta\sigma)$-space were chosen say, ($\sigma_s , ~\Delta\sigma_s$), ($\sigma_m , ~\Delta\sigma_m$) and ($\sigma_l , ~\Delta\sigma_l$) which represent highly-, moderately- and poorly- resolved molecules, respectively. This is as shown in Fig.2B (see, insertion). It may be noted that, mean and variance of the distribution, $G(\sigma , \Delta\sigma)$ need to be predefined and thereafter the process for finding the fortunate molecules needs to be initiated. The molecule that falls in a specific Gaussian distribution (say, $G(\sigma_s , ~\Delta\sigma_s)$ or $G(\sigma_m , ~\Delta\sigma_m)$ or  $G(\sigma_l , ~\Delta\sigma_l)$ ) are designated as fortunate in the respective range ($\sigma, \Delta\sigma$). For example, if the application desires $\sigma = 15~nm$ with a variance of say, $\Delta\sigma = 10~nm$, the fortunate molecules are the ones that falls in the range ($[\sigma-\Delta\sigma, ~\sigma+\Delta\sigma]=[5~nm, 25~nm]$) and obeys Gaussian distribution function $G(\sigma=15~nm, ~\Delta\sigma=10~nm)$. To generate multi-resolution map, more than one Gaussians need to be defined in the search space $(\sigma, ~\Delta\sigma)$. Mathematically, we can define Gaussian as, $ G(\sigma, \Delta\sigma)=A ~\exp{ \biggl[ \frac{-(\sigma^2 -\sigma_0^2)^2}{2\Delta\sigma} \biggr]} $ where, $\sigma_0$ and $\Delta\sigma^2$ are respectively the mean and variance of the function. Here, the mean $\sigma_0$ represent average size of the fortunate molecules whereas, the variance $\Delta\sigma$ represent the size-spectrum of fortunate molecules. \\

The Gaussian probability distribution function ensures that the molecule size (or equivalently LP) that is close to the average size $\sigma_0$ (average LP) are more likely to be considered as fortunate molecules than others that are far from it. Since the distribution theoretically extends from $-\infty$ to $+\infty$, we need to symmetrically fix a lower bound on the probability. We have chosen the corresponding probability as, $P_1 = 0.001 $ which means molecules that have a probability $P > P_1$ are selected and termed as, fortunate molecules. These molecules contribute to the reconstructed image. For multiresolution POSSIBLE microscopy, we defined 3 distinct Gaussian distributions. \\


\subsection{Biological Sample Prepration Protocols}

\subsection{Cell culture}
Standard protocol for culturing NIH3T3 fibroblast cells were followed \cite{smileAPL} \cite{sam2013}. The cells were thawn at low passage and grown overnight with complete growth media (80$\%$ DMEM + 10$\%$ calf bovine serum + $4.5 ~ml$ penicillin streptomycin). Subsequently, the medium was changed to reduce the toxicity caused by freezing medium ($90\%$ complete cell medium + $20\%$ DMSO). A balanced protocol is followed to ensure that density of cells to be about $10^5 / cm^2$ and a maximum confluency of about $80\%$ is ensured for subsequent split followed by transfection with the Dendra2HA-plasmid-DNA. \\

\subsection{Cell Transfection and Fixing}
We have chosen to work with NIH3T3 cells for the proposed study \cite{mikeplos} \cite{matSN} \cite{smileAPL}. NIH3T3 mouse fibroblast cells were transiently transfected with Dendra2-Hemagluttinin (HA) using Lipofectamine 3000 (Life Technologies, Invitrogen) as per the established protocol \cite{mikeplos} \cite{ellens} \cite{sam2013}. The constructs for photoactivable fluorescent protein (Dendra2) tagged with proteins of interest (Hemagglutinin), were plated in a 35 mm disc (with coverslips at the bottom) (Thermofisher Scientific). Cells were grown to the confluency of $\approx ~80\%$ and cleaned with phosphate-buffered saline (PBS). Before proceeding with labeling, cells were checked under transmission light microscope for normal morphology. Subsequently, the cells were fixed with $4\%$ PFA. Post-fixing the cells were sealed with another coverslip using fluorosave solvent (Thermofisher Scientific) for long-time preservation and imaging. The cells were then imaged with high-resolution inverted fluorescence microscope equipped with a broad blue light source (of wavelength in the range, 470-490 nm) and appropriate filter-set is used to observe the green fluorescence (emission maximum, $\lambda =507~nm$) from the transfected cells. This confirms the expression of Dendra2-HA protein in the cell. These cells were subsequently selected and imaged with superresolution microscope (POSSIBLE and SMLM). \\

\subsection{Superresolution Imaging}

Typically, 10,000 frames were recorded at 30 Hz and with an EM gain of 250. The activation (wavelength, $\lambda=405~nm$) and excitation (wavelength, $\lambda=561~nm$) laser powers used at the sample were $112 ~\mu W$ and $6.3~mW$, respectively. A total of 173448 molecules were recorded with an average rate of $\approx 17.3$ molecules per frame. This was followed by particle-size filtering to remove false counting and photon-count filtering to get rid of random background and very weak emitters.  \\

Post fixation to the coverslips, cells were imaged using the super-resolution microscope \cite{smileMRT} \cite{smileAPL}. The molecules thus recorded in several frames were analyzed and localized using developed protocols \cite{cleanAIP}. Major computational tasks involve spot-identification (that indicate emission from a single molecule), thresholding (to eliminate the background) and Gaussian-fit (for identifying single molecules). The analysis determines centroid and variance of the fitted Gaussian giving the location and localization precision of the single molecule. The size of single molecule was determined from its localization precision. Subsequently, single molecule ensembles were subjected to cluster analysis. \\

\subsection{K-means Clustering of HA Molecules and Cluster Properties}
The single molecule analysis of raw data  determines a total of $173448$ molecules for which the positions and localization precision were calculated. A size-based filtering is carried out to filter noisy pixels. This is based on the fact that, less than $3\times 3$ pixels window on the camera chip may not represent single molecule and so the Gaussian-fit to determine its size is not appropriate, whereas larger than $9\times 9$ pixel window may represent two or more closeby single molecules. This is followed by Gaussian filter to filter-out unfortunate molecules. The remaining fortunate molecules belonging to $G(\sigma_l , \Delta\sigma_l)$, $G(\sigma_m , \Delta\sigma_m)$ and $G(\sigma_s , \Delta\sigma_s)$ are $14042$, $18562$ and $14507$, respectively. These molecules were used to reconstruct the respective super-resolved image. Post reconstruction, the clusters were determined by K-means method (inbuilt MATLAB scripts). We identified about 26 clusters for SMLM reconstructed images whereas, 23, 22 and 25 clusters were observed respectively for $G(15~nm , ~10~nm$), G($30~nm, ~2~nm$)$, $ G($50~nm, ~2~nm$). This formed the basis for further analysis such as, cluster area analysis and the number of HA-molecules per cluster (data not shown).  \\

Clustering involves the knowledge of position coordinate of single molecules that lie within the same cluster. We used K-means clustering which is an unsupervised learning method. However it is iterative in nature that begins with an initial guess for the centroids (of the clusters) and the iterations continues untill there is no change to the centroids. K-means use nearest-neighbor method (cityblock) that use sum of absolute differences (L1 distance) for clustering data-points and the centroid is median of points in the cluster. This enables, the determination of important clustering parameters related to influenza infection such as, the cluster area, cluster-density and molecules per cluster. Unlike other implementations, specific shape of the cluster is not assumed for deducing these parameters. This helps in a better estimation of important parameters related to clustering. \\


\section{Discussion}

POSSIBLE superresolution microscopy is a powerful and unique methodology for in-principle infinite resolution {\bf{ of the order of actual single molecule dimension}}. The key that holds the superior resolution is the number of photons emitted by the molecule. This has a direct bearing on the localization-precision ($\Delta_{lp} = \Delta_{psf} / \sqrt{N}$) of the molecule, where $\Delta_{psf}$ is the diffraction-limited PSF and $N$ is the number of emitted photons. It is quite clear that to gain better resolution, the molecule need to be bright and should be capable of emitting a large number of photons. We call these extra-ordinary bright molecules as {\bf{fortunate molecules}}. The brightness of a molecule is the product of quantum efficiency and extinction coefficient (that in turn depend upon transition dipole moment and local environment (solvent) that are free from oxygen species leading to photobleaching) \cite{song}. Thus with the availability of large emitted photons the position of the molecule can be estimated and consequently the LP/size can be reduced. A large collection of fortunate molecules can then be used to reconstruct the map which can be harnessed to understand the biological processes and the underlying mechanisms with unprecedented resolution. \\

The central idea is to identify fortunate molecules and built-up enough statistics to reconstruct {\bf{ultra-superresolved image}}. Such a technique would be able to resolve molecular clusters. Figure 5 demonstrates this by resolving large aggregrates/clusters showing molecular compactness, its density and spread. The ultra-superresolved molecular map generated by POSSIBLE microscopy enables detailed study of molecular clusters which are not possible by traditional localization microscopy (fPALM/PALM). Our study has revealed existence of sub-clusters within large clusters (see, Fig.4), thereby revealing the true nature of HA-assembly post 24 Hrs of transfection. K-means clustering (inbuilt MATLAB scripts) is exploited to determine the aggregration process and calculate cluster size (or equivalently cluster area), HA molecules per cluster and spread-density. These parameters suggest that post 24 Hrs of transfection, clustering happens in small sub-clusters (size $1-2 ~\mu m^2$) before joining together to form large clusters (size $\approx 6 ~\mu m^2$). This is important because small clusters relate to mild infection and drugs can be targeted to prevent large clustering (indication of strong infection). A similar study demonstrates the effect of actin-disrupting drugs on the morphology of HA clusters in Dendra2-HA-transfected NIH3T3 cells \cite{sam2013}.  \\

Another important parameter is the number of HA molecules in clusters which is critical to virion maturation followed by its exit from the infected cell. Here, the traditional SMLM super-resolution microscopy is unable to predict the appearance of these densely-packed clusters, whereas POSSIBLE microscopy shows the presence of dense clusters. In addition, sub-clusters are found in elongated large HA clusters. We attribute this to the ability of POSSIBLE microscopy that can produce better-resolved molecular maps. These observed sub-clusters  range from few tens to few hundred nanometers lying within elongated HA clusters. Earlier studies using fPALM has revealed the presence of elongated clusters \cite{sam2013}.  \\ 

However, proposed methodology has disadvantages too: (1) Since large statistics is sought and detection of fortunate molecules is a rare event so the hunting-time is large, requiring large acquisition time, (2) currently, not many super bright probes are available that can be conjugated with the protein of interest, and (3) POSSIBLE technique is strictly limited to fixed samples. In addition, a meaningful ultra-superresolution map can only be constructed with the availability of sufficient fortunate molecules. These limitations and the difficulty associated with in-vivo study of biological specimens is a handicap for this promising technique. On the bright side, it should be possible to design super bright non-toxic photoactivable probes that can be easily conjugated with the protein-of-interest for studying biological processes \cite{kwon} \cite{banaz}. \\       

Overall, POSSIBLE microscopy gains spatially by sacrificing the temporal resolution and hence it is suitable for fixed-cell imaging. Multiresolution capability of POSSIBLE microscopy suggests that more than one single molecule maps can be reconstructed with used defined resolution. As an example, Fig.2 shows 3 different resolution regimes: highly-resolved ($15\pm 10~nm$), moderately-resolved ($30\pm 2~nm$) and poorly-resolved ($50\pm 2~nm$), respectively. Traditional localization microscopy closely resembles poorly-resolved POSSIBLE microscopy with a large average LP and broad LP-bandwidth. This is further clear from the fPALM images and localization map shown in Fig.3. In addition, proposed technique is readily adaptable and integrable with the existing localization techniques. We envision that POSSIBLE microscopy has the ability to stretch the resolution to single-digit for functional biological imaging.  \\

\section*{Acknowledgment}
The author thanks Prof. Samuel T. Hess, Department of Physics, University of Maine, Orono, USA for fruitful discussion on fPALM superresolution microscopy and support. \\

\section*{Competing interests}
The authors declare no competing interests. \\

\section*{Additional Information}

The programmes, images and videos are available on request. This manuscript incorporates 4 supplementaries.

\end{document}